\documentclass[aps,prl,twocolumn,showpacs]{revtex4}
\usepackage{amssymb}
\usepackage{amsmath}
\usepackage{epsfig}
\usepackage{color}

\newcommand {\cn} {{\rm cn}}

\begin{document}

\title{Stochastic Resonance in the Fermi-Pasta-Ulam Chain}
\author {George Miloshevich${}^{1}$, Ramaz Khomeriki${}^{1,2}$,  Stefano Ruffo${}^3$}
\affiliation {${\ }^{(1)}$ Physics Department, Tbilisi State University, 0128
  Tbilisi (Georgia) \\
${\ }^{(2)}$ Max-Planck-Institut fur Physik komplexer Systeme, 01187 Dresden (Germany)\\
${\ }^{(3)}$ Dipartimento di Energetica ``S. Stecco'' and CSDC, Universit{\`a}
di Firenze, and INFN, via S. Marta, 3, 50139 Firenze (Italy)}

\begin{abstract}
We consider a damped $\beta$-Fermi-Pasta-Ulam chain, driven at one boundary subjected to stochastic
noise. It is shown that, for a fixed driving amplitude and frequency,  increasing the noise intensity, 
the system's energy resonantly responds to the modulating frequency of the forcing signal. Multiple peaks
appear in the signal to noise ratio, signalling the phenomenon of {\it stochastic resonance}.
The presence of multiple peaks is explained by the existence of many stable and metastable states 
that are found when solving this boundary value problem for a semi-continuum approximation of the model. 
Stochastic resonance is shown to be generated by transitions between these states. 
\end{abstract}

\maketitle

Since its theoretical discovery \cite{benzi,nicolis} and experimental verification 
\cite{fauve,mcnamara} {\it stochastic resonance} has become one of the most spread 
topics in many different branches of physics, biophysics and chemistry 
(see Ref. \cite{review} for a review). It has a wide variety of applications, e.g. 
in semiconductors, in neuronal systems, electronic and magnetic systems and, 
recently, even in quantum physics. All these studies are unified by the common idea 
to represent a bistable system as particle motion in a double well potential. According 
to this simplification the system begins to jump from one well to the other, following 
the periodic forcing, when a resonant noise level is attained. These concepts have
wide applications to optical systems. 
In the ring laser, bistability is due to the left-right propagation symmetry of the
system, while in absorptive optical media it is a consequence
of the existence of different output powers corresponding to the same input pump 
intensity \cite{arimondo}. This latter situation has some similarities with the
system we discuss in this Letter.

This approach cannot be straightforwardly applied to spatially
extended systems, where different stable and metastable excited
regimes can coexist, expecially when the forcing is applied locally.
As discussed in \cite{longtin}, the presence of metastable states does
not hinder stochastic resonance. In this Letter, we show that stochastic
resonance can be realized for the well-known Fermi-Pasta-Ulam (FPU) 
chain \cite{fpu} of anharmonic oscillators.
Up to now stochastic resonance in coupled one dimensional systems
has been considered only when each element of the
chain displays bistable properties \cite{lindner,lindner1}. 
In our case, bistability characterizes the whole chain, being originated by the coexistence of
different stationary regimes corresponding to a single driving amplitude. 
This coexistence is also at the basis of the nonlinear bistability effect \cite{ramaz11,RJ2,lepri} 
which appears when the chain is harmonically driven
at one end with out-band frequency and is due to presence of
breather like excitations \cite{Thierry} in the system.

The equations of motion for the $\beta$-FPU chain (interparticle potential 
with quadratic and quartic terms) are
\begin{figure}[ht]
\epsfig{file=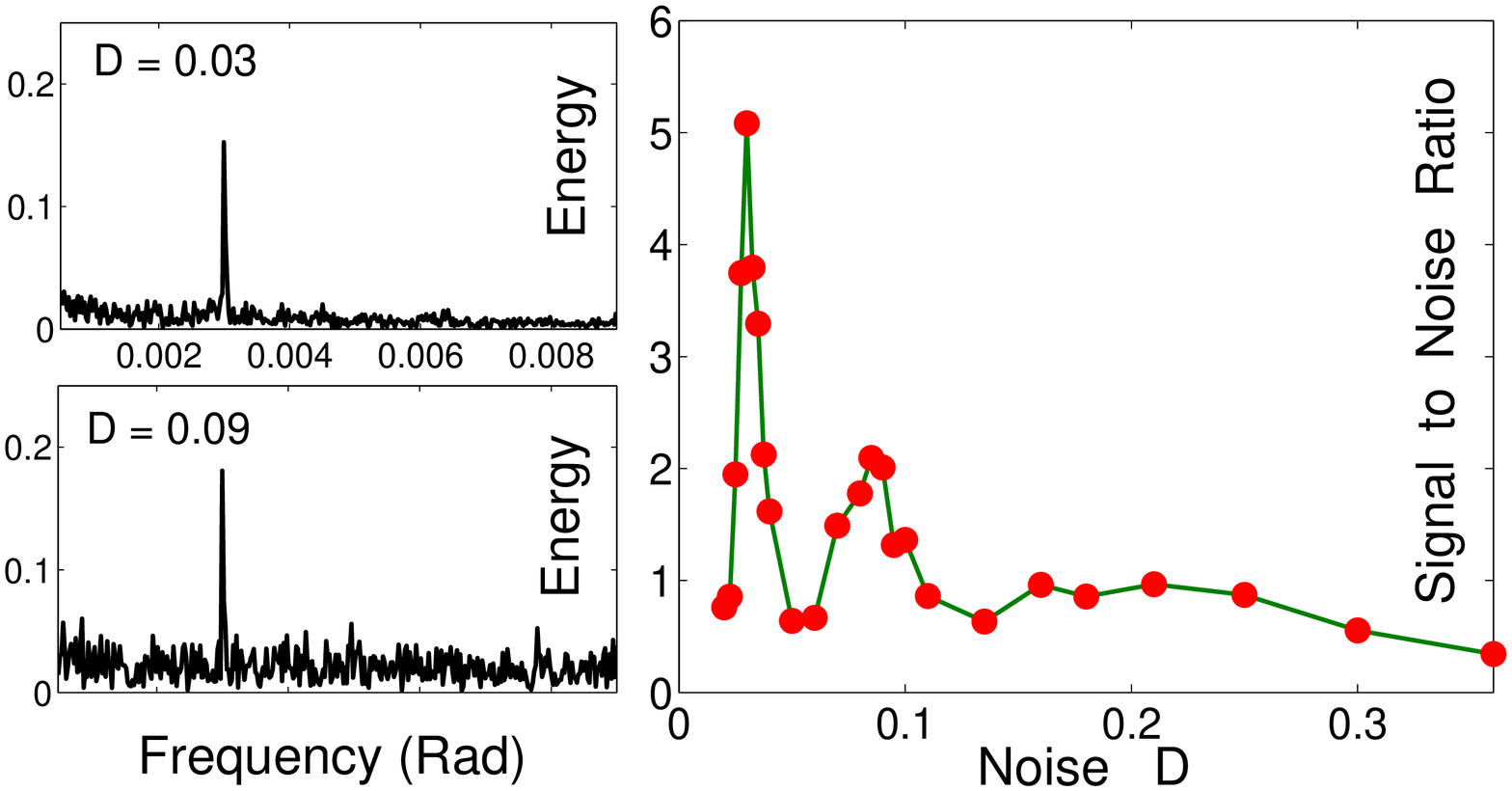,width=0.9\linewidth} 
\epsfig{file=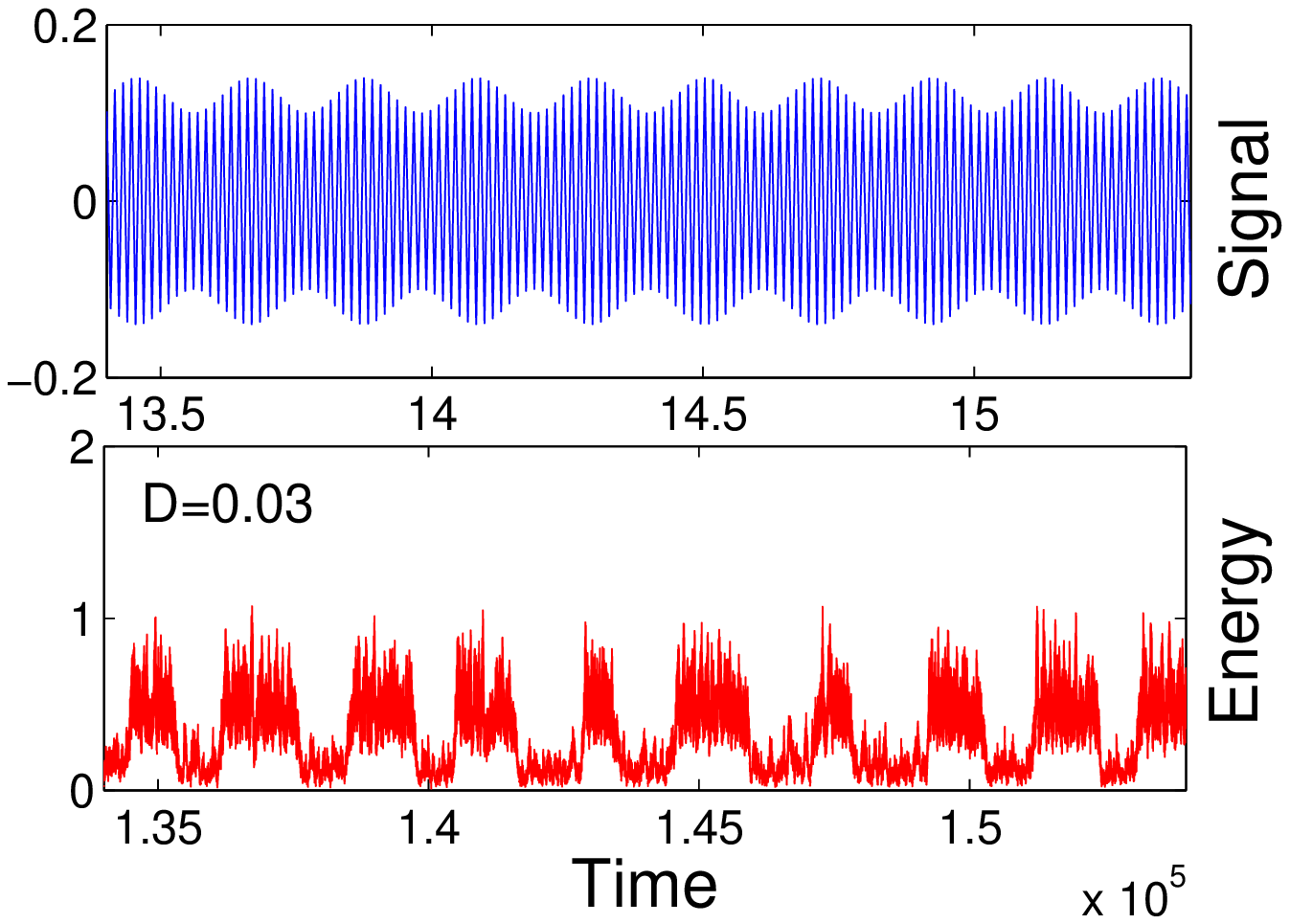,width=0.85\linewidth}
\caption{Typical signatures of stochastic resonance. In the top-left graph we display the power 
spectrum of the energy signal of the FPU chain for the two noise strengths $D=0.03$ 
and $D=0.09$ at which the signal to noise ratio (SNR) shows the marked peaks reported in the top-right
graph. The forcing signal, with carrier frequency $\Omega=2.05$ and modulation $W=0.003$,
is plotted in the central graph. The time evolution of the energy for the noise intensity
corresponding to the first peak of the SNR plot is shown in the bottom graph: it
displays the characteristic bistable features of stochastic resonance.}
\end{figure}
\begin{eqnarray}
\ddot u_n &=&u_{n+1}+u_{n-1}-2u_n+(u_{n+1}-u_n)^3 \nonumber \\ &&
+(u_{n-1}-u_n)^3-\sigma \dot u_n+\xi_n(t),
\label{FPU}
\end{eqnarray}
where $u_n$ stands for the displacement of the $n$-th unit mass of the chain in dimensionless units
($n=1,2\ldots, N$); $\sigma$ is the damping parameter (fixed throughout the paper to $\sigma=0.04$)
and $\xi_n(t)$ is a zero average Gaussian white noise applied independently to each oscillator 
with the following autocorrelation function
\begin{equation}
\langle \xi_m(t)\xi_n(0)\rangle=2 D \delta(t)\delta_{mn}.
\label{correl}
\end{equation}
We force the system at the left end as follows
\begin{equation}
u_0(t)=A \cos (Wt) \cos (\Omega t),
\label{drive}
\end{equation}
while the right boundary condition is free
\begin{equation}
u_N(t)=u_{N+1} (t)~.
\label{free}
\end{equation}
As commented below, the value of $N$ should not be too large, in all simulations
presented here $N=6$.

In Fig. 1 we display several characteristic features of stochastic resonance in the FPU chain.
Let us first concentrate the attention on the top-right plot, where we show the signal to noise
ratio in decibels. Those who are familiar with stochastic resonance may be surprised to observe
multiple peaks in this plot while, usually, a bell shaped curve with a single maximum is found.
We will explain below this observation, providing an analytical argument based on the continuum
limit of the FPU model. The presence of multiple peaks derives from the existence of multiple
stable and metastable states for a given forcing amplitute $A$ (the equivalent of having multiple
wells in the potential). The forcing signal (\ref{drive}), shown in the central graph in Fig. 1, 
has a carrier frequency $\Omega$ which is chosen slightly above the upper edge of the linear spectrum 
$\Omega>\omega_0=2$ in order to excite breather type localisations, and is modulated with a much
smaller frequency $W \ll \Omega$. This latter frequency plays the role of the forcing frequency appearing
in standard stochastic resonance. There is a wide range of values of both $A$ and $W$ for
which we can observe stochastic resonance in FPU, given that we respect the important condition
$\sigma \gg W$. It must be however remarked that the amplitude $A$ should be in a range where 
a stable stationary state of the chain, discussed below, coexists with a metastable one. 
When this condition is met, we vary the noise strength $D$ until we observe the typical oscillations
of the energy of the system shown in the lower graph of Fig. 1. The power Fourier spectrum of the
energy signal shows a sharp peak at the driving frequency $W$ when the noise intensity is at
resonance (top-left graph in Fig. 1).

We develop below our analytical approach. Let us begin by solving the boundary value 
problem for the FPU chain in the absence of damping and noise.
It is well known that the FPU chain is characterized by the following linear dispersion relation 
$\omega(k) =\sqrt{2(1-\cos k)}$, hence the spectrum has band edge frequency $\omega_0=2$. 
In order to derive stationary weakly nonlinear solutions (see e.g. Ref. \cite{flach}) of 
Eq. \eqref{FPU},  we seek for solutions of the following standard form \cite{koslepri,ramaz}:
\begin{equation}
u_n(t)=\frac{(-1)^n}{2}\left[e^{2it}\phi(n,t)+c.c\right],
\label{phi}
\end{equation}
where $c.c.$ denotes the complex conjugated term and $\phi(n,t)$ varies slowly with respect 
to both its arguments $n$ and $t$. These solutions should be considered as modulations of upper 
band edge oscillations. The boundary value problem \eqref{drive} can be rewritten 
for $\phi(n,t)$ in the form
\begin{equation}
\phi(0,t)=A e^{i(\Omega-2)t} \qquad \phi(N+1/2,t)=0,
\label{drive1}
\end{equation}
where the second condition derives from $\phi(N+1)=-\phi(N)$. Moreover,
the condition $\delta \omega= \Omega-2 \ll 1$ must be satisfied for the function $\phi(n,t)$
to be slowly varying with time.

Substituting then \eqref{phi} into Eqs.~\eqref{FPU}, and assuming $\phi(n,t)$ to be a continuous 
function of its variables, one gets, in the weakly nonlinear limit and neglecting higher order derivatives in 
$n$ and $t$, the following nonlinear Schr\"odinger equation \cite{koslepri,ramaz}:
\begin{equation}
i\frac{\partial \phi}{\partial t}+\frac{1}{4}\frac{\partial^2 \phi}{\partial n^2}+3|\phi|^2\phi=0.
\label{schroedinger}
\end{equation}
We now assume that the system synchronizes with the boundary \eqref{drive1} and define a new parameter $B$
\begin{equation}
\phi(n,t)=e^{i t\delta \omega}\varphi(n), \qquad B=\frac{\partial \varphi(n)}{\partial n}\biggr|_{ n=N+1/2}~,
\label{drive2}
\end{equation}
(where $\varphi(n)$ is a real function of $n$) and, as a result, we get the following relation
\begin{equation}
\left(\frac{\partial \varphi}{\partial n}\right)^2=B^2+4(\Omega-2)\varphi^2-6\varphi^4
\label{elliptic1}
\end{equation}
\begin{figure}[ht]
\epsfig{file=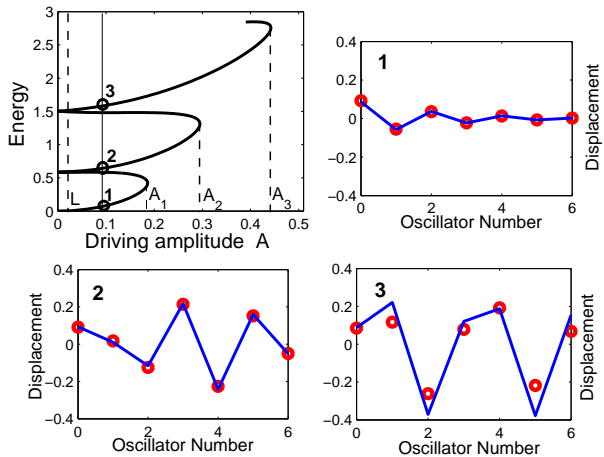,width=0.95\linewidth}
\caption{In the top-left plot the energy of the approximate solutions \eqref{elliptic2} of the 
FPU chain obtained using our semi-continuum approach is displayed as a function of the driving amplitude $A$.
The intersections of the vertical solid line at $A=0.09$ with the different branches of the curve 
correspond to multiple solutions of the consistency relation \eqref{elliptic3}. Type 1 ($E_1\simeq 0.1$) and 
2 ($E_2\simeq 0.6$) solutions, which turn out to be stable, are compared with numerical simulations (points) 
in the top-right and bottom-left panels. A type 3 ($E_3\simeq 1.6$) solution, which is metastable (see text), 
is represented in the bottom-right panel.
$A_1=0.18$, $A_2=0.31$ and $A_3=0.44$ are threshold values for the excitation of the different solutions: $A_1$ is the driving 
amplitude for which the transition from a type 1 stationary state to a type 2 stationary state 
occurs, while $A_2$ marks the transition from type 2 to type 3 (metastable) solutions, and so on.
In the figure, we also display a feature which is present because of the damping and is not taken
into account by our approximate solutions: an additional lower threshold L is present (vertical dashed line), below 
which all the solutions collapse to the type 1 lowest energy state.}
\end{figure}
and its solution in terms of Jacobi elliptic functions \cite{book,RJ2}, which, after substitution into 
Eqs. \eqref{drive2} and \eqref{phi} gives the following approximate solution for the relative 
displacements of the oscillators
\begin{equation}
u_n=(-1)^nQ\cos(\Omega t)\,\cn\left[2\sqrt{\gamma}\left(N+\frac{1}{2}-n\right)-{\cal K}(k),\, k\right]
\label{elliptic2}
\end{equation}
where ${\cal K}(k)$ is the complete elliptic integral of the first kind with a modulus $k$ and all 
the constant are defined via the single free parameter $B$ as follows
\begin{equation}
\gamma^2=\delta\omega^2+\frac{3}{2}B^2; \quad Q^2=\frac{\delta\omega+\gamma}{3}; 
\quad k^2=\frac{3B^2}{4\gamma(\gamma-\delta\omega)}~.
\label{parameters}
\end{equation}
\begin{figure}[t]
\epsfig{file=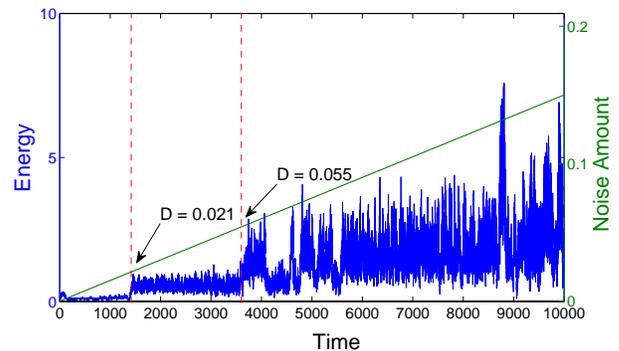,width=\linewidth}
\caption{Time evolution of the energy of the FPU chain for a constant driving
amplitude $A=0.18$ and a linearly varying noise intensity. Transition from a type 1 to a type 2
state is shown to occur when the noise intensity reaches the threshold value $D=0.021$, while
the transition to a type 3 states takes place at $D=0.055$. In this simulation, damping is
present and the driving is not modulated, $W=0$.}
\end{figure}
The free parameter $B$ is defined by the condition of adaptation to the boundary \eqref{drive}, 
leading to the following consistency relation
\begin{equation}
A=Q\,\cn\left[2\sqrt{\gamma}\left(N+1/2\right)-{\cal K}(k),\, k\right].
\label{elliptic3}
\end{equation}
From this equation one gets all compatible values of $B$ for a fixed driving 
amplitude $A$. It turns out that there are multiple values of $B$ that solve
Eq. (\ref{elliptic3}), which in turn implies the existence of multiple
compatible patterns of the type (\ref{elliptic2}), which have correspondingly 
different energies.
In Fig. 2 we display these energies as a function of the driving amplitude
$A$. Three branches of solutions are present. For instance, considering
the forcing amplitude $A=0.09$, five possible energy values are found: among
these two are unstable, two are stable and one is metastable, as described
in the caption of Fig. 2. For the type 1 and 2 solutions in Fig. 2, one has an 
excellent correspondence between the analytical expression \eqref{elliptic2} 
and the direct numerical simulations of the FPU chain. For type 3 solutions the 
approximation is not as good and, besides that, the solution is metastable: in
numerical simulations it destabilizes and restabilizes during time evolution.
We do not display in Fig. 2 solutions corresponding to higher energies, because
the semi-continuouum approximation fails and one should use truncated 
wave approximations \cite{kosprl}.

Summarizing, the system is characterized by several threshold amplitudes. When the 
driving amplitude exceeds the first threshold $A_1$, the type 1 stable state jumps 
into the type 2 stable state. By further increasing the driving up to the second 
threshold $A_2$, the transition to the type 3 metastable state occurs. 
In the presence of damping, when reducing the amplitude of the driving, the system
does not remain in the higher energy states. Indeed, a lower threshold amplitude 
appears, $L=0.03$, below which the system goes back to the lowest energy stable
state. Afterwards, when we will consider stochastic resonance, we will modulate the 
driving signal in such a way that its amplitude remains above $L$ and below
$A_1$ (typically $A \in [0.10,0.14]$), in order to realize stochastic transitions 
between all states.
Furthermore, we note that for a fixed driving amplitude $A$ the stable and metastable
states are characterized by sufficiently far separated energies. The presence 
of these ``energy levels" can be monitored by keeping the driving amplitude constant 
and changing the noise intensity, as shown in Fig. 3. 
It is clearly seen that for the fixed boundary driving amplitude $A=0.18$ the transition 
between type 1 and type 2 stable states occurs at the noise intensity  $D=0.021$, while at 
$D=0.055$ the system goes from a type 2 stable state to a type 3 
metastable state. One should also observe that the averaged energies of each state 
well correspond to the ones derived from the approximate analytical solutions \eqref{elliptic2}.
\begin{figure}[t]
\epsfig{file=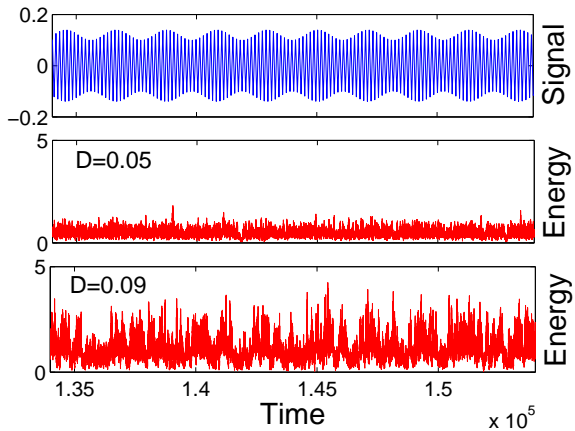,width=0.95\linewidth}
\caption{Time evolution of the signal (top) and system's energy (central and bottom) at the noise 
intensities $D=0.05$ (corresponding to the point between first two peaks of the SNR in Fig. 1)
and $D=0.09$ (second peak of the SNR), respectively.}
\end{figure}

Now we are ready to discuss the main result of this paper. Simulating the FPU 
model \eqref{FPU} with the driving signal seed frequency $\Omega=2.05$ and modulating
frequency $W=0.003$, such that the driving signal amplitude $A$ varies in the range $[0.10,0.14]$, 
in the presence of both damping and noise, we monitor the time evolution of the 
system's energy for increasing noise intensities $D$, and then analyse the properties
of its power spectrum $P(\omega)$. The signal to noise ratio (SNR) is defined as 
$SNR=10 \log_{10} (P(\omega_{peak})/N(\omega_{peak}))$, where 
$\omega_{peak}$ is the frequency at which the power spectrum displays a clear peak
(see the top-left plots in Fig. 1) and $N(\omega_{peak})$ is the power of the background noise at
the same frequency.
For low noise intensities there is no transition between the different states and the system's 
energy oscillates around $E_1$. For the noise intensity value $D=0.03$ (bottom graph in Fig. 1)
one observes an oscillation between the state with energy $E_1=0.1$ and that with energy $E_2=0.6$. 
Increasing further noise intensity, the SNR drops and the energy fluctuates around the value $E_2=0.6$
(middle plot in Fig. 4 for $D=0.05)$). Finally, for the noise intensity $D=0.09$, which corresponds 
to the second peak of the SNR in Fig. 1, we observe noisy transitions between the metastable state
of energy $E_3=1.6$ and two stable states with energies $E_1$ and $E_2$ (see 
the bottom graph of Fig. 4). We do not discuss here the smooth third peak observed in Fig. 1, which
should be connected to transitions with higher energy states, of which we do not possess
a clear analytical characterization.

As remarked above, we have performed all our numerical analysis for {\it short} FPU chains
(throughout the paper $N=6$), but the observation of this phenomenology is not
restricted to this case. Indeed, it should be mentioned that, if in the equations for the $\beta$-FPU chain 
\eqref{FPU} we would have rescaled the strength of the nonlinear coupling to $f$, 
the semi-continuum approach developed here would have produced exactly the same solutions \eqref{elliptic2},
provided that the number of particles $N$ would have been increased maintaining the ratio $\sqrt{f}/N$ constant.

Concluding, we have considered a damped $\beta$-FPU chain forced at one boundary with a modulated
signal under the action of an increasing noise level. For specific values of the noise intensity, the
power spectrum of the system's energy displays sharp peaks at the modulating frequency. Correspondingly,
also the signal to noise ratio shows pronounced peaks as a function of noise intensity. This is a clear 
observation of {\it stochastic resonance} in an extended system. In contrast to all previous studies, 
multiple peaks of the SNR are observed, instead of a single broad bump. In order to explain this feature, we have developed a
semi-continuum analytical approach, which allows to point out the presence of multiple stable
and metastable solutions for a given forcing amplitude. Stochastic resonance is shown to be 
generated by transition between these states. These studies could be readily extended to realistic 
physical systems with local forcing.

\paragraph{Acknowledgements.} R. Kh. acknowledges financial support of the 
Georgian National Science Foundation (Grant No GNSF/STO7/4-197) and the USA Civilian 
Research and Development Foundation (award No GEP2-2848-TB-06). This work is part
of the PRIN07 project {\it Statistical physics of strongly correlated systems at and
out of equilibrium: exact results and quantum field theory methods}.

\end{document}